\documentclass{article}
\usepackage{slashed,amsmath,amssymb,enumitem}
\usepackage[papersize={8.5in,11in}]{geometry}
\usepackage{graphicx}
\begin{document}
\title{Quantum Gravity and the Cosmological Constant Problem}
\author{J. W. Moffat\\~\\
Perimeter Institute for Theoretical Physics, Waterloo, Ontario N2L 2Y5, Canada\\
and\\
Department of Physics and Astronomy, University of Waterloo, Waterloo,\\
Ontario N2L 3G1, Canada}
\maketitle
\begin{abstract}
 A finite and unitary nonlocal formulation of quantum gravity is applied to the cosmological constant problem. The entire functions in momentum space at the graviton-standard model particle
 loop vertices generate an exponential suppression of the vacuum density and the cosmological constant to produce agreement with their observational bounds.
\end{abstract}	

%\begin{fmffile}{ewunifigs}

\section{Introduction}

A nonlocal quantum field theory and quantum gravity theory has been formulated that leads to a finite, unitary and locally gauge invariant theory~\cite{Moffat,Moffat2,Moffat3,Woodard,Kleppe,Cornish,Cornish2,Hand,Woodard2,Clayton,Paris,Troost,Joglekar,Moffat4}. For quantum gravity the finiteness of
quantum loops avoids the problem of the non-renormalizabilty of local quantum gravity~\cite{Veltman,Sagnotti}.

The finiteness of the nonlocal quantum field theory draws from the fact that factors of $\exp[{\cal
K}(p^2)/\Lambda^2]$ are attached to propagators which suppress any
ultraviolet divergences in Euclidean momentum space, where $\Lambda$ is an
energy scale factor. An important feature of the field theory is {\it that only the
quantum loop graphs have nonlocal properties}; the classical tree graph
theory retains full causal and local behavior.

Consider first the 4-dimensional spacetime to be approximately flat Minkowski
spacetime. Let us denote by $f$ a generic local field and write
the standard local Lagrangian as
\begin{equation}
{\cal L}[f]={\cal L}_F[f]+{\cal L}_I[f],
\end{equation}
where ${\cal L}_F$ and ${\cal L}_I$ denote the free part and the interaction part
of the action, respectively, and
\begin{equation}
{\cal L}_F[f]=\frac{1}{2}f_i{\cal K}_{ij}f_j.
\end{equation}
In a gauge theory the action would be the Becchi, Rouet, Stora, Tyutin (BRST)
gauge-fixed action including ghost fields in the invariant action required
to fix the gauge~\cite{Becchi,Tyutin}. The kinetic operator ${\cal K}$ is fixed by
defining a Lorentz-invariant distribution operator:
\begin{equation}
\label{distribution}
{\cal E}\equiv \exp\biggl(\frac{{\cal K}}{2\Lambda^2}\biggr)
\end{equation} and the operator:
\begin{equation}
{\cal O}=\frac{{\cal E}^2-1}{{\cal
K}}=\int_0^1\frac{d\tau}{\Lambda^2}\exp\biggl(\tau\frac{{\cal K}}{\Lambda^2}\biggr).
\end{equation}

The regularized interaction Lagrangian takes the form
\begin{equation}
{\hat {\cal L}}_{I}=-\sum_n(-g)^nf{\cal I}[{\cal F}^n,{\cal O}^{(n-1)})]f,
\end{equation}
where $g$ is a coupling constant and ${\cal F}$ is a vertex function form factor. The
decomposition of ${\cal I}$ in order $n=2$ is such that the operator ${\cal O}$
splits into two parts ${\cal F}^2/{\cal K}$ and $-1/{\cal K}$. For Compton amplitudes
the first such term cancels the contribution from the corresponding lower order
channel, while the second term is just the usual local field theory result for that
channel. The action is then invariant under an extended nonlocal gauge
transformation. The precise results for QED were described in ref.~\cite{Moffat2}.

The regularized action is found by expanding
${\hat{\cal L}}_I$ in an infinite series of interaction terms. Since ${\cal F}$ and
${\cal O}$ are entire function of ${\cal K}$ the higher interactions are also
entire functions of ${\cal K}$. This is important for preserving the Cutkosky rules
and unitarity, for an entire function does not possess any singularities in the
finite complex momentum plane.

The Feynman rules are obtained as follows: Every leg of a diagram is connected to a local propagator,
\begin{equation}
\label{regpropagator}
D(p^2)=\frac{i}{{\cal K}(p^2)+i\epsilon}
\end{equation}
and every vertex has a form factor ${\cal F}^k(p^2)$, where $p$ is the momentum
attached to the propagator $D(p^2)$, which has the form
\begin{equation}
{\cal F}^k(p^2)\equiv{\cal E}^k(p^2)=\exp\biggl(\frac{\cal K}{2\Lambda_k}\biggr),
\end{equation}
where $k$ denotes the particle nature of the external leg in the vertex.
The formalism is set up in Minkowski spacetime and loop integrals are formally
defined in Euclidean space by performing a Wick rotation. This facilitates the
analytic continuation; the whole formalism could from the outset be developed in
Euclidean space.

We will demonstrate how the nonlocal transcendental entire function in momentum space that generates the finite and unitary standard model (SM) and
quantum gravity (QG) loops to all orders of perturbation theory, produces an exponential suppression of the estimated very large vacuum
density and cosmological constant in local quantum field theory. This can solve the severe fine-tuning cosmological constant problem, avoiding a naturalness problem and the need for
an anthropic and multiverse solution.

\section{Nonlocal Quantum Gravity}

We expand the metric around a smooth fixed background spacetime:
\begin{equation}
\label{background}
g_{\mu\nu}={\bar g}_{\mu\nu}+h_{\mu\nu}.
\end{equation}
By restricting ourselves to perturbation theory and a fixed geometrical background, we lose
general covariance (diffeomorphism invariance). However, we still maintain gauge invariance of the gravitational calculations under
the gauge group of the fixed background metric, e.g., for a fixed Minkowski metric background the action is invariant under local
Poincar\'{e} transformations, while for a de Sitter background metric the action will be invariant under the group of de Sitter
transformations. Although we lose general covariance in our perturbation calculations of gravitational
scattering amplitudes, the basic physical properties such as finiteness of loop amplitudes, gauge invariance and unitarity will
be expected to lead to correct and reliable physical conclusions. For the sake of simplicity, we shall only consider expansions about
Minkowski spacetime.

Let us define ${\bf g}^{\mu\nu}=\sqrt{-g}g^{\mu\nu}$, where ${\bf g}=
{\rm det}({\bf g}^{\mu\nu})$ and $\partial_\rho{\bf g}={\bf
g}_{\alpha\beta}\partial_\rho{\bf g}^{\alpha\beta}{\bf g}$. We
can then write the local gravitational action $S_{\rm grav}$ in
the form~\cite{Goldberg}:
\begin{equation}
\label{action}
S_{\rm grav}=\int d^4x{\cal L}_{\rm grav}=\frac{1}{2\kappa^2}\int
d^4x [({\bf g}^{\rho\sigma}{\bf g}_{\lambda\mu} {\bf
g}_{\kappa\nu}
$$ $$
-\frac{1}{2}{\bf g}^{\rho\sigma} {\bf
g}_{\mu\kappa}{\bf g}_{\lambda\nu}
-2\delta^\sigma_\kappa\delta^\rho_\lambda{\bf
g}_{\mu\nu})\partial_\rho{\bf g}^{\mu\kappa} \partial_\sigma{\bf
g}^{\lambda\nu}
$$ $$
-\frac{2}{\alpha}\partial_\mu{\bf
g}^{\mu\nu}\partial_\kappa{\bf g}^{\kappa\lambda}
\eta_{\nu\lambda}
+{\bar C}^\nu\partial^\mu X_{\mu\nu\lambda}C^\lambda],
\end{equation}
where $\kappa^2=32\pi G$ and we have added a gauge fixing term with the parameter $\alpha$, $C^\mu$ is the Fadeev-Popov ghost field and
$X_{\mu\nu\lambda}$ is a differential operator:
\begin{equation}
X_{\mu\nu\lambda}=\kappa(-\partial_\lambda\gamma_{\mu\nu}
+2\eta_{(\mu\lambda}\gamma_{\kappa\nu)}\partial^\kappa)
+(\eta_{(\mu\lambda}\partial_{\nu)}-\eta_{\mu\nu}\partial_\lambda).
\end{equation}

We expand the local interpolating graviton field ${\bf g}^{\mu\nu}$ as
\begin{equation}
{\bf g}^{\mu\nu}=\eta^{\mu\nu}+\kappa\gamma^{\mu\nu}+O(\kappa^2).
\end{equation} Then,
\begin{equation}
{\bf g}_{\mu\nu}=\eta_{\mu\nu}-\kappa\gamma_{\mu\nu}
+\kappa^2{\gamma_\mu}^\alpha{\gamma_\alpha}_\nu+O(\kappa^3).
\end{equation}

The gravitational Lagrangian density is expanded as
\begin{equation}
{\cal L}_{\rm grav}={\cal L}^{(0)}+\kappa{\cal L}^{(1)}
+\kappa^2{\cal L}^{(2)}+....
\end{equation}
In the limit $\alpha\rightarrow\infty$, the Lagrangian density
${\cal L}_{\rm grav}$ is invariant under the gauge
transformation
\begin{equation}
\delta\gamma_{\mu\nu}=X_{\mu\nu\lambda}\xi^\lambda,
\end{equation}
where $\xi^\lambda$ is an infinitesimal vector quantity.

To implement nonlocal quantum gravity, we introduce the ``stripping'' graviton propagator in the gauge $\alpha=-1$:
\begin{equation}
{\tilde D}_{\alpha\beta\mu\nu}(p)
=\frac{1}{2}(\eta_{\alpha\mu}\eta_{\beta\nu}+\eta_{\alpha\nu}\eta_{\beta\mu}
-\eta_{\alpha\beta}\eta_{\mu\nu}){\cal O}_0(p),
\end{equation}
while the ghost stripping propagator is given by
\begin{equation}
{\tilde D}^{\rm ghost}_{\mu\nu}(p)=\eta_{\mu\nu}{\cal O}_0(p),
\end{equation}
where
\begin{equation}
{\cal O}_0(p)=\frac{{\cal E}^2_0-1}{p^2}.
\end{equation}
We choose ${\cal E}_0^2=\exp(-p^2/\Lambda_G^2)$ and we see that the local propagator can be obtained from the nonlocal
propagator minus the stripping propagator
\begin{equation}
\frac{1}{p^2}=\frac{\exp(-p^2/\Lambda_G^2)}{p^2}-{\cal O}_0(p).
\end{equation}
The stripping propagators are used to guarantee that the tree-level
graviton-graviton scattering amplitudes are identical to the local,
point-like tree-level amplitudes, which couple only to physical gravitons.

The graviton propagator in the fixed de Donder gauge $\alpha=-1$~\cite{Donder} in momentum space is given by
\begin{equation}
D_{\mu\nu\rho\sigma}(p)
=\frac{\eta_{\mu\rho}\eta_{\nu\sigma}+\eta_{\mu\sigma}\eta_{\nu\rho}
-\eta_{\mu\nu}\eta_{\rho\sigma}}{p^2+i\epsilon},
\end{equation}
while the graviton ghost propagator in momentum space is
\begin{equation}
D^{\rm ghost}_{\mu\nu}(p)=\frac{\eta_{\mu\nu}}{p^2+i\epsilon}.
\end{equation}

The on-shell vertex functions are unaltered from their local antecedents, while virtual
particles are attached to nonlocal vertex function form factors. This destroys the
gauge invariance of e.g. graviton-graviton scattering and requires an iteratively
defined series of ``stripping'' vertices to ensure the decoupling of all unphysical
modes. Moreover, the local gauge transformations have to be extended to nonlinear,
nonlocal gauge transformations to guarantee the over-all invariance of the
regularized amplitudes. The quantum gravity perturbation theory
is invariant under generalized, nonlinear field representation dependent transformations, and it is
finite to all orders. At the tree graph level all unphysical polarization states are decoupled and nonlocal effects
will only occur in graviton and graviton-matter loop graphs. Because the gravitational tree graphs are purely local there is a
well-defined classical GR limit. The finite quantum gravity theory is well-defined in four real spacetime dimensions or in any higher
D-dimensional spacetime.

We quantize by means of the path integral operation
\begin{equation}
\langle 0\vert T^*(O[{\bf g}])\vert 0\rangle_{\cal E}=\int[D{\bf g}]
\mu[{\bf g}]({\rm gauge\, fixing})
O[{\bf g}]\exp(i\hat S_{\rm grav}[{\bf g}]).
\end{equation}
The quantization is carried out
in the functional formalism by finding a measure factor
$\mu[{\bf g}]$ to make $[D{\bf g}]$ invariant under the
classical symmetry. Because we have extended the gauge symmetry to nonlinear,
nonlocal transformations, we must also supplement the
quantization procedure with an invariant measure
\begin{equation}
{\cal M}=\Delta({\bf g}, {\bar C}, C)D[{\bf
g}_{\mu\nu}]D[{\bar C}_\lambda]D[C_\sigma]
\end{equation}
such that $\delta {\cal M}=0$.

\section{The Cosmological Constant Problem}

The cosmological constant problem is considered to be the most
severe hierarchy problem in modern physics~\cite{Weinberg,Polchinski,Martin,Burgess}.

We can define an effective cosmological constant
\begin{equation}
\lambda_{\rm eff}=\lambda_0+\lambda_{\rm vac},
\end{equation}
where $\lambda_0$ is the `bare' cosmological
constant in Einstein's classical field equations,
and $\lambda_{\rm vac}$ is the contribution that arises from the
vacuum density $\lambda_{\rm vac}=8\pi G\rho_{\rm vac}$. The observational bound on $\rho_{\rm vac}$ is
\begin{equation}
\label{vacbound}
\rho_{\rm vac} \leq 10^{-47}\, ({\rm GeV})^4,
\end{equation}
corresponding to the the bound on $\lambda_{\rm vac}$:
\begin{equation}
\label{lambdabound}
\lambda_{\rm vac} \leq 10^{-84}\,{\rm GeV}^2.
\end{equation}

Zeldovich~\cite{Zeldovich} showed that the zero-point
vacuum fluctuations must have a Lorentz invariant form
\begin{equation}
T_{{\rm vac}\,\mu\nu}=\lambda_{\rm vac}g_{\mu\nu},
\end{equation}
consistent with the equation of state $\rho_{\rm vac}=-p_{\rm
vac}$. Thus, the vacuum within the framework of particle quantum
physics has properties identical to the cosmological constant. In
quantum theory, the second quantization of a classical field of
mass $m$, treated as an ensemble of oscillators each with a
frequency $\omega(k)$, leads to a zero-point energy
$E_0=\sum_k\frac{1}{2}\hbar\omega(k)$. An evaluation of the
vacuum density obtained from a summation of the zero-point energy
modes gives
\begin{equation}
\rho_{\rm vac}
=\frac{1}{(2\pi)^2}\int_0^{M_c}dkk^2(k^2+m^2)^{1/2}
\sim\frac{M^4_c}{16\pi^2},
\end{equation}
where $M_c$ is the cutoff. Taking $M_c\sim M_{\rm Planck}\sim 10^{19}$ GeV, we get $\rho_{\rm vac}\sim 122$ orders of
magnitude greater than the observed value.
Already at the level of the standard model, we get $\rho_{\rm vac}\sim (10^2\,{\rm GeV})^4$ which is
$55$ orders of magnitude larger than the bound (\ref{vacbound}).
To agree with the experimental bound (\ref{vacbound}), we would
have to invoke a very finely tuned cancellation of $\lambda_{\rm
vac}$ with the `bare' cosmological constant $\lambda_0$, which is
generally conceded to be theoretically unacceptable.

We adopt a model consisting of a photon field $A_\mu$ coupled to gravity.  We have for the effective field Lagrangian density:
\begin{equation}
\label{LA}
{\cal L}_A=-\frac{1}{4}(-{\bf g})^{-1/2}{\bf g}^{\mu\nu}{\bf g}^{\alpha\beta}F_{\mu\alpha}F_{\nu\beta},
\end{equation}
where
\begin{equation}
F_{\mu\nu}=\partial_\nu A_\mu-\partial_\mu A_\nu.
\end{equation}

We have
\begin{equation}
{\cal L}_A^{(0)}=-\frac{1}{4}\eta^{\mu\nu}\eta^{\alpha\beta}F_{\mu\alpha}F_{\nu\beta},
\end{equation}
and
\begin{equation}
{\cal L}_A^{(1)}=-\frac{1}{4}\biggl(\eta^{\mu\nu}\gamma^{\alpha\beta}+\eta^{\alpha\beta}\gamma^{\mu\nu}
-\frac{1}{2}\eta^{\mu\nu}\eta^{\alpha\beta}\gamma\biggr)F_{\mu\alpha}F_{\nu\beta}.
\end{equation}
We include in the Lagrangian density ${\cal L}_A^{(0)}$ an additional gauge-fixing piece $-\frac{1}{2}(\partial^\mu A_\mu)^2$. For a particular gauge no Faddeev-Popov ghost particles and diagrams contribute to the lowest order photon-graviton self-energy calculation. The local photon propagator has the form
\begin{equation}
D^{\rm A}_{\mu\nu}(p)=\frac{\eta_{\mu\nu}}{p^2+i\epsilon}.
\end{equation}

The graviton-A-A vertex in momentum space is given by
\begin{align}
{\cal V}_{\alpha\beta\lambda\sigma}(q_1,q_2)
=\eta_{\lambda\sigma}
q_{1(\alpha}q_{2\beta)}-\eta_{\sigma(\beta}q_{1\alpha)}q_{2\lambda}
-\eta_{\lambda(\alpha}q_{1_\sigma}q_{2\beta)}\nonumber\\
+\eta_{\sigma(\beta}\eta_{\alpha)\lambda}q_1{\cdot q_2}
-\frac{1}{2}\eta_{\alpha\beta}(\eta_{\lambda\sigma}
q_1{\cdot q_2}-q_{1\sigma}q_{2\lambda}),
\end{align}
where $q_1,q_2$ denote the momenta of the two $Vs$ connected to the graviton with momentum $p$.

The lowest order correction to the graviton vacuum loop will have the form
\begin{align}
\label{PolV}
\Pi^{\rm GA}_{\mu\nu\rho\sigma}(p)
=-\kappa^2\exp(-p^2/\Lambda_G^2)\int d^4q
{\cal V}_{\mu\nu\lambda\alpha}(p,q){\cal F}(q^2)
D^{A\,\lambda\delta}(q)\nonumber\\
\times{\cal V}_{\rho\sigma\kappa\delta}(p,q-p){\cal F}((q-p)^2)
D^{A\,\alpha\kappa}(q-p).
\end{align}
Let us adopt the entire functions ${\cal F}(p^2)_{\rm SM}=\exp(-p^2/2\Lambda_{\rm SM}^2)$ and ${\cal F}(p^2)_=\exp(-p^2/2\Lambda_G^2)$ in Euclidean momentum space, scaled by the SM energy scale $\Lambda_{\rm SM}$ and the gravitational energy scale $\Lambda_G$, respectively. We obtain
\begin{align}
\label{Ptensor}
\Pi^{\rm GV}_{\mu\nu\rho\sigma}(p)=-\kappa^2\exp(-p^2/\Lambda_G^2)
\int\frac{d^4q\eta^{\lambda\delta}\eta^{\alpha\kappa}}{q^2(q-p)^2}{\cal V}_{\mu\nu\lambda\alpha}(p,q)\nonumber\\
\times{\cal V}_{\rho\sigma\kappa\delta}(p,q-p)\exp\biggl(-q^2/2\Lambda^2_{\rm SM}\biggr)
\exp\biggl(-(q-p)^2/2\Lambda^2_{\rm SM}\biggr).
\end{align}
As usual, we must add to (\ref{Ptensor}) the contributions from the tadpole vector-graviton diagrams and the invariant measure diagram.

We observe that from power counting of the momenta in the integral (\ref{Ptensor}), we obtain
\begin{align}
\label{VacPol}
\Pi^{\rm GA}_{\mu\nu\rho\sigma}(p)\sim
\kappa^2\exp(-p^2/\Lambda_G^2)N_{\mu\nu\rho\sigma}(\Lambda_{\rm SM},p^2),
\end{align}
where $N_{\mu\nu\rho\sigma}(\Lambda_{\rm SM},p^2)$ is a finite contribution to $\Pi^{\rm GA}_{\mu\nu\rho\sigma}(p)$.  ${{{\Pi^{\rm GA}_\mu}^\mu}^\sigma}_\sigma(p)$ vanishes at $p^2=0$, as it should because of gauge invariance to this order and the massless graviton.

The vector field vertex form factor, {\it when coupled to SM gauge bosons}, will have the form
\begin{equation} {\cal E}^{\rm SM}(p^2)
=\exp\biggl(-p^2/2\Lambda_{SM}^2\biggr).
\end{equation}
If we choose $\Lambda_{SM}\gtrsim 1$ TeV, then we will reproduce the low energy SM experimental results and ${\cal F}^{\rm SM}(p^2)$ becomes ${\cal F}^{\rm SM}(0)=1$ on the mass shell $p^2=0$~\cite{Moffat,Moffat2}.

\section{Cosmological Constant Problem and Quantum Gravity}

The cosmological constant problem is considered to be the most severe hierarchy problem in modern physics~\cite{Weinberg,Polchinski,Martin,Burgess}. Can our quantum gravity theory solve the cosmological constant problem? The cosmological constant is a non-derivative coupling in the Lagrangian density ${\cal L}_{\rm grav}$:
\begin{equation}
\label{lambda}
{\cal L}_\lambda=-\frac{4}{\kappa^2}\lambda\sqrt{-g}.
\end{equation}
In diagrammatic terms, it is a sum of zero momentum and zero temperature vacuum fluctuation loops coupled to external gravitons. The problem is to explain why the magnitude of $\lambda$ is suppressed to be zero or a very small value when compared to observation.

Let us initially consider the basic lowest order vacuum
fluctuation diagram computed from the matrix element in flat Minkowski spacetime:
\begin{equation}
\rho_{\rm vac}\sim\rho^{(2)}_{\rm vac}\sim
g^2\int d^4pd^4p'd^4k\delta(k+p-p')\delta(k+p-p')
$$ $$
\times\frac{1}{k^2+m^2}{\rm Tr}\biggl(\frac{i\gamma^\sigma
p_\sigma-m_f}{p^2+m_f^2}\gamma^\mu\frac{i\gamma^\sigma
p'_\sigma-m_f}{p^{'2}+m_f^2}\gamma_\mu\biggl)
\exp\biggl[-\biggl(\frac{p^2+m_f^2}{2\Lambda^2_{SM}}\biggr)
-\biggl(\frac{p'^2+m^2_f}{2\Lambda^2_{SM}}\biggr)
-\biggl(\frac{k^2+m^2}{2\Lambda^2_{SM}}\biggr)\biggr],
\end{equation}
where $g$ is a coupling constant associated with the standard model.
We have considered a closed loop made of a SM fermion of mass $m_f$, an anti-fermion
of the same mass and an internal SM boson propagator of mass $m$; the scale
$\Lambda_{\rm SM}\sim 1$ TeV. This leads to the result
\begin{equation}
\rho_{\rm vac}\sim\rho^{(2)}_{\rm vac}\sim 16\pi^4g^2\delta^4(a)\int_0^{\infty}dpp^3\int_0^{\infty}dp'p^{'3}
\biggl[\frac{-P^2+p^2+p^{'2}+4m_f^2}{(P+a)(P-a)}\biggr]
\frac{1}{(p^2+m_f^2)(p'^2+m_f^2)}
$$ $$
\times \exp\biggl[-\frac{(p^2+p'^2+2m^2_f)}
{2\Lambda^2_{SM}}-\frac{P^2+m^2}{2\Lambda^2_{SM}}\biggr],
\end{equation}
where $P=p-p'$ and $a$ is an infinitesimal constant which formally regularizes the infinite volume factor
$\delta^4(0)$. We see that $\rho_{\rm vac}\sim \Lambda_{\rm SM}^4$. By choosing our nonlocal energy scale for the standard model, $\Lambda_{\rm NL}\sim\Lambda_{\rm SM}\sim 1\, {\rm TeV}=10^3\,{\rm GeV}$, we have reduced the magnitude of the vacuum density by 64 orders of magnitude compared to having $\Lambda_{\rm SM}\sim \Lambda_{\rm Planck}\sim 10^{19}\,{\rm GeV}$.

In Minkowski spacetime, the sum of all {\it disconnected}
vacuum diagrams is a constant factor $C$ in the
scattering S-matrix $S'=SC$. Since the S-matrix is unitary
$\vert S'\vert^2=1$, then we must conclude that $\vert
C\vert^2=1$, and all the disconnected vacuum graphs can be
ignored. This result is also known to follow from the Wick ordering of the field
operators. However, due to the equivalence principle {\it gravity couples to all
forms of energy}, including the vacuum energy density $\rho_{\rm vac}$, so we can no
longer ignore these virtual quantum fluctuations in the presence of a non-zero
gravitational field.

\begin{figure}[t]
\begin{center}{\includegraphics[width=\linewidth]{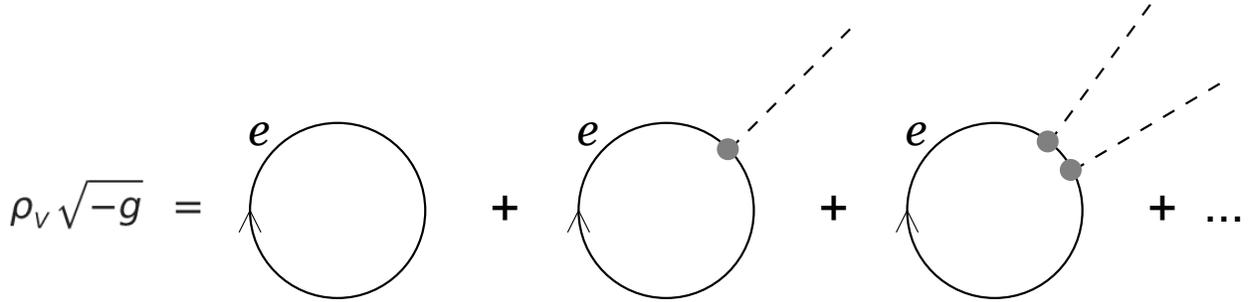}}\end{center}
\caption{Electron vacuum fluctuation loops coupled to gravitons generating a vacuum density.\label{fig:selfEnergyGrav.eps}}
\end{figure}

We can view the cosmological constant as a non-derivative coupling
of the form $\lambda_0\sqrt{-g}$ in the Einstein-Hilbert action (See Fig. 1). Quantum corrections to $\lambda_0$ come from
loops formed from massive SM states, coupled to external graviton lines at
essentially zero momentum. The massive SM states are far off-shell. Experimental
tests of the standard model involving gravitational couplings to the SM states are
very close to being on-shell. Important quantum corrections to $\lambda_0$ are
generated by a huge extrapolation to a region in which gravitons couple to SM
particles which are far off-shell.

To reduce the size of the vacuum density to agree with the observational bound, we must discover how gravity can couple to the vacuum energy
density and generate an exponential damping of the very large $\rho_{\rm vacSM}$. This exponential suppression of $\rho_{\rm vacSM}$ can
be produced by nonlocal QG. There will be virtual graviton legs connected to the quantum gravity-standard model loops by a nonlocal vertex entire function,
$\exp(-p_G^2/2\Lambda_G^2)$. We see from (\ref{VacPol}) that the standard model vacuum polarization and vacuum density are reduced by the nonlocal graviton vertex interaction:
\begin{equation}
\rho_{\rm vac}\sim \exp(-{\bar p_G}^2/2\Lambda_G^2)\rho_{\rm vacSM},
\end{equation}
where ${\bar p_G}=\langle p_G^2\rangle^{1/2}$ is an average mean of the virtual graviton momentum $p_G$. If we choose ${\bar p_G}=16.49\Lambda_G$, then we have
\begin{equation}
\label{VacSupp}
\rho_{\rm vac}\sim \exp(-{\bar p_G}^2/2\Lambda_G^2)\rho_{\rm vacSM}\sim 10^{-47}\,{\rm GeV}^4,
\end{equation}
and we have reduced the cosmological constant contribution, $\lambda_{\rm vac}=8\pi G\rho_{\rm vac}$, to the observed bound $\lambda_{\rm vacObs}\leq 10^{-84}\,{\rm GeV}^2$, where we have used the nonlocal energy scale $\Lambda_{SM}\sim 1$ TeV in the coupling to standard model particles. The size of $\Lambda_G$ should be small enough to allow for soft graviton momenta. This can be achieved by choosing $\Lambda_G < 1$ MeV, so that the mean virtual graviton momentum ${\bar p}_G=16.5\Lambda_g < 17$ MeV. The size of the exponential suppression of the vacuum energy in (\ref{VacSupp}) can be related to a violation of the weak equivalence principle through the electrostatic energy associated with the vacuum polarization of atomic hydrogen coupled to external gravitons~\cite{Polchinski,Martin}, so the choice of $\Lambda_G$ can play an important role. However, the violation of the equivalence principle can be affected by the material environment, namely, the difference between the atomic matter environment versus the vacuum energy density in empty space at extra-galactic distance scales.

\section{Conclusions}

The nonlocal formulation of quantum gravity provides a finite, unitary and locally gauge invariant perturbation theory. The vertex functions associated with point-like interactions in local quantum field theory are replaced by smeared out nonlocal vertex functions controlled by transcendental entire functions. The choice of entire function in momentum space $\exp(-p^2/2\Lambda^2)$, where $\Lambda=\Lambda_{\rm SM}\sim 1$ TeV and $\Lambda=\Lambda_G$ for the standard model and quantum gravity, respectively, guarantees the finiteness of all quantum loops. We have demonstrated how the vacuum fluctuations involving SM loops can be exponentially dampened by the entire functions for the graviton-standard model particle vertex functions. For a mean value of the virtual graviton momenta the exponential suppression can reduce the vacuum density fluctuations and the cosmological constant to agree with the cosmological observational bounds.

\section*{Acknowledgements}
The John Templeton Foundation is thanked for its generous support of
this research. The research was also supported by the Perimeter
Institute for Theoretical Physics. Research at the Perimeter Institute for Theoretical Physics is supported by the Government of Canada through industry Canada and by the Province of Ontario through the Ministry of Research and Innovation (MRI). I thank Martin Green and Viktor Toth for helpful discussions.
%\end{fmffile}

\end{document}